# Prediction of Magnetoelectric behavior in $Bi_2MnTiO_6$

Lokanath Patra[a] and P. Ravindran[a,b,*]

[a]*Central University of Tamil Nadu, Thiruvarur, 610101, India*
[b]*Center for Materials Science and Nanotechnology and Department of Chemistry, University of Oslo, Box 1033, Blindern, N-0315, Oslo, Norway*
*raviphy@cutn.ac.in

**Abstract.** We present results from *ab initio* calculations based on density functional theory for bismuth-based double perovskite $Bi_2MnTiO_6$. Using total energy calculation with stress and force minimization we have predicted the equilibrium crystal structure for $Bi_2MnTiO_6$ considering potential structures into the calculation. We have predicted that the ground state of $Bi_2MnTiO_6$ will be a noncentrosymmetric rhombohedral structure with space group R3c. Our spin polarized calculation for different possible collinear magnetic configurations we found that $Bi_2MnTiO_6$ will be an insulator with *G*-type antiferromagnetic ordering in its ground state. The coexistence of both stereochemically active Bi-6*s* lone pair and the $Ti^{4+}$ with $d^0$-ness which bring covalency results in the stabilization of noncentrosymmetric structure and thus ferroelectricity. Our orbital projected density of states plot shows that the $Mn^{2+}$ in $Bi_2MnTiO_6$ will be at high spin state with a spin moment of 4.28 $\mu_B$. Hence $Bi_2MnTiO_6$ is predicted to be a magnetoelectric material.

## INTRODUCTION

There has been growing interest in multiferroic materials due to the coupled magnetic, electric and/or structural order parameters which result in (anti)ferromagnetic, (anti)ferroelectric and/or ferroelastic properties in the same phase. Due to these combinations, the magnetic properties can be manipulated by applying either electric or magnetic field.[1] Bi based multiferroics have become a center of attraction due to high value of electric polarization[2,3] and $Bi^{3+}$ is more environment friendly than $Pb^{2+}$. In addition, the stereochemical activity of $Bi^{3+}$ is being exploited in magnetic oxides, with the goal of forming ferromagnetic-ferroelectric coupling. $BiMnO_3$ undergoes monoclinic to monoclinic transition at 474 K and monoclinic to orthorhombic transition at 770 K[4–6]. Montaneri *et al*[7] found that below 770 K $BiMnO_3$ has a centrosymmetric *C2/c* space group by neutron powder diffraction experiments. Since the *C2/c* structure of $BiMnO_3$ has inversion symmetry, the net electric polarization is zero. The mechanism of breaking the inversion symmetry with magnetism was considered recently in ref. 8[8]. It was found that the peculiar orbital ordering was realized below 474 K gives rise to ferromagnetic interactions between nearest-neighbor spins which compete with longer-range antiferromagnetic interaction. Solovyev and Pchelkina[9] predicted that *C2/c* symmetry in $BiMnO_3$ is spontaneously broken by hidden AFM interactions, which lowers the actual symmetry to the *Cc* space group which allows for the existence of both ferromagnetic order and ferroelectric polarization. The Ti doping in $BiMnO_3$ is expected to increase the magnetoelectric coupling in $Bi_2MnTiO_6$ as the $d^0$-ness of $Ti^{4+}$ will give rise to off center displacement to the system. However, there is no theoretical prediction has been done for this compound with Ti doping up to now. This motivated the present study.

## COMPUTATIONAL DETAILS

The present calculations were performed using the Vienna ab-initio simulation package (VASP)[10,11] within the projected augmented plane wave (PAW)[12] method together with the Perdew-Burke-Ernzerhof generalized gradient approximation (GGA)[13] for the exchange correlation potential. A very large basis set of 800 eV for the plane wave cut off was used in order to reproduce the structural parameters correctly[2]. Structural optimizations were continued until the forces on the atoms had converged to less than 1 meV/Å and the pressure on the cell had minimized within the constraint of constant volume. 6x6x6 **k**- mesh was used for *R3* structure and similar k-point density has been used for all the other potential structures considered here. In order to identify the correct magnetic ground state, we have considered non-magnetic (NM), ferromagnetic (FM) and three antiferromagnetic (AFM) configurations such as A, C, G-AF configurations.[14] The Born effective charges (BEC) are calculated for the ground state configuration using the so called ''Berry phase finite difference approach'' in which the electronic contribution to the change of polarization is estimated using modern theory of polarization[15].

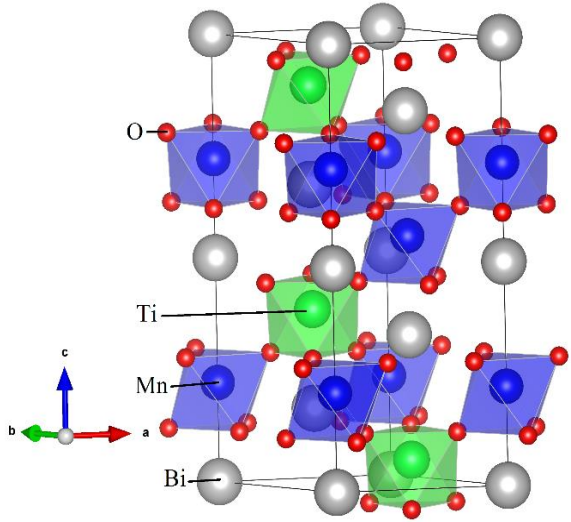

**FIGURE 1.** Crystal structure of Bi$_2$MnTiO$_6$ in the *R3* structure. Highly distorted MnO$_6$ and TiO$_6$ octahedras are corner shared through oxygen. Also note that Bi atoms are not present in the middle between the octahedras owing to the displacement of Bi due to the presence of 6$s$ lone pair electrons.

## RESULTS AND DISCUSSIONS

For the study of structural phase stability, we have considered six closely related potential structure types: tetragonal *Amm2*, cubic *Fm-3m*, tetragonal *P4mm*, rhombohedral *R3r*, monoclinic C121, and monoclinic *C2/c* (Fig. 2). The structures were fully relaxed for all volumes considered in the present study using force as well as stress minimization. From Fig. 2 it is clear that the noncentrosymmteric rhombohedral *R3* (space group 146) structure is energetically lower among the structural configurations considered here. The calculated volume vs total energy curve was fitted to Murnaghan equation of states to get equilibrium volume, bulk modulus and its pressure derivative. The optimized equilibrium structural parameters are as follows: Volume ($V$) = 129.70 Å$^3$, $a$ = 5.69 Å, α = 59.80 and the positions of the atoms are Bi (1a) (0.000/0.507), Mn/Ti (1a) (0.233/0.729), O1 (3a) (0.568, 0.926, 0.388) and O2 (3a) (0.882, 0.454, 0.037). The equilibrium bulk modulus and its pressure derivative obtained from the fitting are 80.7 GPa and 6.34 respectively. It may be noted that the equilibrium structure obtained for Bi$_2$MnTiO$_6$ is closely related to the *R3c* structure found experimentally and theoretically for $BiFeO_3$[2,16] and to the $R3r$ structure predicted theoretically for $Bi_2FeTiO_6$[17].

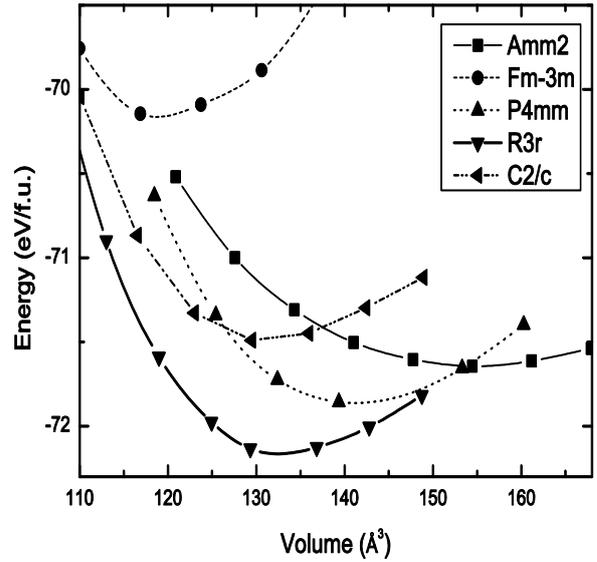

**FIGURE 2.** Calculated cell volume vs total energy for antiferromagnetic Bi$_2$MnTiO$_6$ in different structural arrangements (structure types[#] being labeled on the illustration).

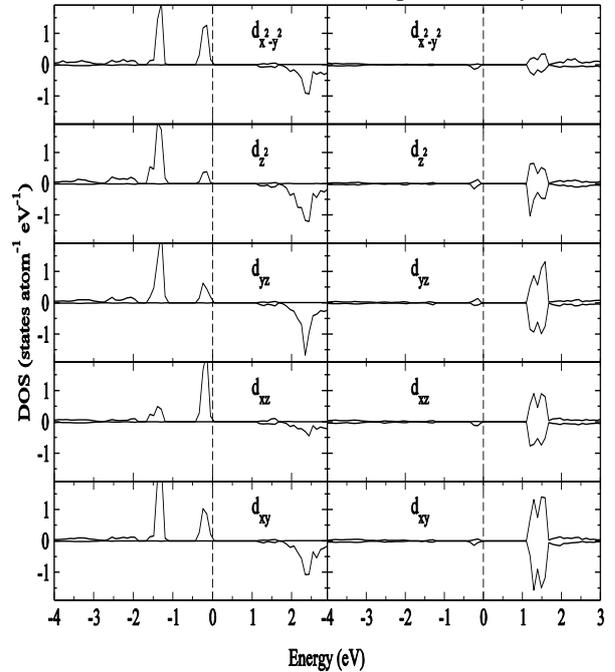

**FIGURE 3.** Calculated orbital-projected density of states of Mn$^{2+}$ (left) and Ti$^{4+}$ (right) for Bi$_2$MnTiO$_6$ in the ground state *G*-AFM configuration.

---

[#]The curve for *C121* is not shown here because its energy is very high as compared to the energy for the ground state configuration *R3*

**TABLE 1.** Calculated Born effective charge for antiferromagnetic $Bi_2MnTiO_6$

| $Z^*$ | Position | xx | yy | zz | xy | xz | yx | yz | zx | zy |
|---|---|---|---|---|---|---|---|---|---|---|
| Bi | 1a | 5.177 | 4.999 | 4.380 | -0.505 | -0.241 | 0.049 | -0.688 | -0.573 | 4.380 |
| Bi | 1a | 4.565 | 4.592 | 4.670 | 0.127 | 0.004 | -0.062 | 0.129 | 0.105 | 4.670 |
| Mn | 1a | 2.886 | 2.788 | 2.500 | -0.048 | -0.212 | -0.167 | -0.240 | -0.152 | 2.500 |
| Ti | 1a | 5.593 | 5.482 | 5.128 | 0.204 | -0.495 | -0.505 | -0.193 | -0.045 | 5.128 |
| O | 3b | -2.742 | -3.243 | -2.692 | -0.635 | -0.141 | -0.409 | 0.422 | -0.532 | -2.692 |
| O | 3b | -3.115 | -2.831 | -2.681 | -0.329 | 0.746 | 0.008 | -0.402 | 0.576 | -2.681 |

Our calculations show that the magnetic ground state of $Bi_2MnTiO_6$ will be $G$-AFM. The calculated total energies with respect to ground state G-AF configuration for ferromagnetic, $A$-AFM, $C$-AFM and nonmagnetic states are 7 meV, 477 meV, 476 meV and 1851 meV/f.u., respectively. The integrated spin density at the Mn site is found to be 4.28 $\mu_B$/Mn atom. This clearly indicates that Mn2+ ion in $Bi_2MnTiO_6$ in HS state. In an ideal octahedral crystal field, the $d$-levels will split into triply degenerate $t_{2g}$ ($d_{xy}$, $d_{xz}$, $d_{yz}$) and doubly degenerate $e_g$ ($d_{x^2-y^2}$, $d_{z^2}$) levels. In pure ionic picture $Mn^{2+}$ with five d electrons will fill these energy levels with a low spin (LS) state ($t_{2g}^5 e_g^0$) or high spin (HS) state ($t_{2g}^3 e_g^2$) having spin moments 1 $\mu_B$ and 5 $\mu_B$, respectively. Our orbital projected DOS in Fig. 3 for the $G$-AFM configuration shows HS state for $Mn^{3+}$ and hence only majority spin channels are filled five $d$ orbitals. The Ti atom tends to form an oxidation state of $Ti^{4+}$, and hence our orbital projected DOS show nearly empty d orbital at the Ti site.

The formal valence of Bi, Mn, Ti and O in $Bi_2MnTiO_6$ are 3+, 2+, 4+ and 2-, respectively. When we include AFM ordering, the space group changes from $R3r$ to $R3$. Owing to the site symmetry, the diagonal components of $Z^*$ are anisotropic for all the values of $Z^*$ for all the ions in Table 1. The noticeable off-diagonal components in the oxygen sites confirm the presence of co-valent bonding between O 2p and transition metal 3d orbitals. It is generally expected that there is considerable covalent bonding between Mn-O and Ti-O within $MnO_6$ and $TiO_6$ octahedra, respectively which results as the additional dynamical charge with respect to the well-known ionic values. As the ions listed in the table are having high values of BEC, $Bi_2MnTiO_6$ is expected to have greater polarization than $BiMnO_3$.

## CONCLUSION

The electrical polarization of $BiMnO_3$ can be improved by doping foreign transition metal ions. As $d^0$ ness brings extra contribution to polarization we have substituted Mn ion with $Ti^{4+}$ ion that resulted magnetolelctric $Bi_2MnTiO_6$. $Mn^{2+}$ is in HS state and also our BEC shows strong covalent interaction between transition metals and oxygen. The stabilization of ferroelectric phase is originating from the lone pair electron at the $Bi^{3+}$ site and covalency from $Ti^{4+}$ ion.

## ACKNOWLEDGMENTS

The authors are grateful to the Research Council of Norway for providing computing time at the Norwegian supercomputer facilities. This research was supported by the Indo-Norwegian Cooperative Program (INCP) via Grant No. F. No. 58-12/2014(IC).

#The curve for *C121* is not shown here because its energy is very high as compared to the energy for the ground state configuration *R3*